\documentclass[traditabstract]{aa} 
\usepackage{graphicx}
\def\la{\lower.5ex\hbox{$\; \buildrel < \over \sim \;$}}
\def\ga{\lower.5ex\hbox{$\; \buildrel > \over \sim \;$}}
\begin{document}
   \title{The influence of the cluster environment on the large-scale radio continuum emission of 8 Virgo cluster spirals}


   \author{B.~Vollmer\inst{1}, M.~Soida\inst{2}, A.~Chung\inst{3}, R.~Beck\inst{4},
     M.~Urbanik\inst{2}, K.T.~Chy\.zy\inst{2}, K.~Otmianowska-Mazur\inst{2}, \and J.H.~van Gorkom\inst{5}}

   \institute{CDS, Observatoire astronomique de Strasbourg, 11, rue de l'universit\'e,
          67000 Strasbourg, France \and
          Astronomical Observatory, Jagiellonian University,
          Krak\'ow, Poland \and
          Smithsonian Astrophysical Observatory, 60 Garden Street, Cambridge, MA 02138, USA \and
          Max-Planck-Institut f\"{u}r Radioastronomie, Auf dem H\"{u}gel 69, 53121 Bonn, Germany \and
          Department of Astronomy, Columbia University, 538 West 120th Street, New York,
          NY 10027, USA}

   \date{Received ; accepted }


  \abstract
{The influence of the environment on the polarized and total power
radio continuum emission of cluster spiral galaxies is investigated.
We present deep scaled array VLA 20 and 6~cm observations including polarization
of 8 Virgo spiral galaxies. These data are combined with existing
optical, H{\sc i}, and H$\alpha$ data. 
Ram pressure compression leads to sharp edges of the total power distribution
at one side of the galactic disk. These edges coincide with H{\sc i} edges.
In edge-on galaxies the extraplanar radio emission can extend further than the H{\sc i}
emission. 
In the same galaxies asymmetric gradients in the degree of polarization give additional
information on the ram pressure wind direction. The local total power
emission is not sensitive to the effects of ram
pressure. The radio continuum spectrum might flatten in the compressed region only
for very strong ram pressure.
This implies that neither the local star formation rate
nor the turbulent small-scale magnetic field are significantly
affected by ram pressure. Ram pressure compression
occurs mainly on large scales ($\ga 1$~kpc) and is primarily detectable in
polarized radio continuum emission.
}


   \authorrunning{Vollmer et al.}
   \titlerunning{Large-scale radio continuum emission of 8 Virgo cluster spirals}

   \maketitle
%

\section{Introduction\label{sec:introduction}}

Radio continuum emission is due to relativistic electrons of density
$N_{\rm e}$ gyrating around the interstellar magnetic field $B$: $S
\propto N_{\rm e}B_{\perp}^2$, where $S$ is the intensity of synchrotron emission
and $B_{\perp}$ is the component of the total magnetic field in the sky plane. 
The galactic magnetic field can be divided
into a small-scale and large-scale component compared to the
resolution of the radio continuum observations, which is typically
about 1~kpc in nearby galaxies. The small-scale magnetic field is
due to turbulent gas motions and is therefore tangled. The
large-scale magnetic field is due to a galactic dynamo. 
Polarized emission is due to the regularly
oriented, large-scale magnetic field, but it can also be caused by an
alignment of anisotropic small-scale magnetic fields, 
produced by stretching and compression of
small-scale magnetic structures. While the large-scale 
unidirectional fields yield a non-zero Faraday rotation, this is not
the case for aligned anisotropic small-scale fields.
The small-scale magnetic field is typically a factor of 2--5 larger
than the regular large-scale magnetic field in spiral arms and 1--2
times larger in the interarm regions (Beck 2001). Whenever there is
enhanced turbulence due to enhanced star formation, the small-scale
magnetic field is increased and the large-scale magnetic field is
diminished (see, e.g., Beck 2007).

Deep VLA observations at $\lambda$6~cm have shown that the
distribution of polarized radio continuum emission of Virgo cluster
spiral galaxies is strongly asymmetric, with elongated ridges
located in the outer galactic disk (Vollmer et al. 2004, Chy\.zy et
al. 2006, Chy\.zy et al. 2007, Vollmer et al. 2007). These features
are not found in similar observations of field galaxies, where the
distribution of $\lambda$6~cm polarized emission is generally
relatively symmetric and strongest in the interarm regions (Beck
2005). The polarized radio continuum emission is sensitive to
compression and shear motions within the galactic disks occurring
during the interaction between the galaxy and its cluster
environment. These interactions can be of tidal nature (with the cluster
potential: Byrd \& Valtonen 1990; Valluri 1993, rapid flybys of
massive galaxies, galaxy ``harassment'': Moore et al. 1998) or
hydrodynamic nature (ram pressure stripping: Gunn \& Gott 1972).

On the other hand, total radio continuum emission is sensitive to
star formation which gives rise to the radio--FIR correlation
(see, e.g. Murphy et al. 2008).
Based on 1.4~GHz radio continuum observations Gavazzi et al. (1991) 
showed that this correlation that is
shared by spiral galaxies in a huge luminosity interval, is different
for cluster and isolated galaxies: the cluster galaxies have an increased
radio/FIR ratio. A similar increase has been observed at 4.85~GHz 
and 10.55~GHz (Niklas et al. 1995) for 6 
out of 45 observed Virgo spirals including NGC~4388 and NGC~4438.
Gavazzi \& Boselli (1999) studied the
distribution of the radio/NIR luminosity (RLF) of late-type galaxies
in 5 nearby galaxy clusters. They found that the RLF of Cancer,
ACO~262 and Virgo are consistent with that of isolated galaxies.
However, galaxies in ACO~1367 and Coma have their radio emissivity
enhanced by a factor $\sim 5$ with respect to isolated objects.
Multiple systems in the Coma cluster bridge also show an enhanced
radio/NIR emission. Gavazzi \& Boselli (1999) argue that the latter
effect is due to increased star formation caused by tidal
interactions, whereas the enhanced radio/NIR ratio in ACO~1367
and Coma is due to ram pressure compression of the magnetic field.
Murphy et al. (2009) investigated the radio--FIR relation of Virgo
cluster galaxies. They showed that ram pressure affected galaxies
have global radio flux densities that are enhanced 
by a factor of $2$--$3$ compared to isolated galaxies. 

The typical spectral slope of synchrotron emission from spiral galaxies
between 1.4~GHz and 4.85~GHz is $S \propto \nu^{-0.8}$
(Gioia et al. 1982, Klein 1990). V\"{o}lk \& Xu (1994) proposed that a shock-induced 
reacceleration of relativistic electrons during ram pressure compression might lead to 
an enhanced total power emission and a flattening of the radio continuum spectrum.
However, based on integrated galaxy properties,
Niklas (1995) and Vollmer et al. (2004) do not find any significant difference
between the mean spectral index of Virgo cluster and field spiral galaxies.

It is still an open question if environmental
interactions in a galaxy cluster can locally enhance the total power
radio continuum emission of a galaxy and if they can alter the
spectral index as suggested by Vollmer et al. (2004) for NGC~4522 in
the Virgo cluster.

In this article we present the total power radio continuum
observations of 8 Virgo spiral galaxies observed by Vollmer et al. (2007)
in polarization. 
These galaxies were carefully selected based on the following criteria:
(i) they show signs of interaction with the cluster environment
such as tidal interactions and/or ram pressure stripping,
(ii) VIVA H{\sc i} data (Chung et al. 2009) are available, and (iii) their 6cm total
power emission is strong. In previous work we elaborated interaction
scenarios for most of the galaxies in our sample.

Deep VLA 20 and 6~cm observations including
polarization are presented in Sect.~\ref{sec:observations}. 
In Sect.~\ref{sec:results} we present for each galaxy the
(i) 6~cm total power emission distribution superposed on DSS B band image together with the apparent B vectors,
(ii) 20~cm total power emission distribution on a DSS B band image together with the apparent B vectors,
(iii) 6~cm polarized emission distribution on the H{\sc i} distribution,
(iv) H$\alpha$ emission distribution (from Goldmine, Gavazzi et al. 2003) on the spectral index map, and 
(v) 6~cm polarized emission distribution on the degree of polarization.
The galaxy properties at different wavelengths are compared to each other in Sect.~\ref{sec:comparison}.
We discuss our results in
Sect.~\ref{sec:discussion} and give our conclusions in Sect.~\ref{sec:conclusions}.

\section{Observations\label{sec:observations}}

The 8 Virgo spiral galaxies were observed at 4.85~GHz between November 8, 2005
and January 10, 2006 with the Very Large Array (VLA) of the National
Radio Astronomy Observatory (NRAO)\footnote{NRAO is a facility of
National Science Foundation operated under cooperative agreement by
Associated Universities, Inc.} in the D array configuration. The
band passes were $2\times 50$~MHz. We used 3C286 as the flux
calibrator and 1254+116 as the phase calibrator, the latter of which
was observed every 40~min. Maps were made for both wavelengths using
the AIPS task IMAGR with ROBUST=3. The final cleaned maps were
convolved to a beam size of $18'' \times 18''$. The bright radio
source M~87 caused sidelobe effects enhancing the rms noise level of
NGC~4438. 
In addition, we observed the 8 galaxies at 1.4~GHz on August 15, 2005
in the C array configuration.
The band passes were $2\times 50$~MHz. We used the same calibrators as
for the 4.85~GHz observations. The final cleaned maps were
convolved to a beam size of $20'' \times 20''$.
The rms levels of the 20 and 6~cm total power and
polarized intensity data are shown in Table~\ref{tab:table}.
We obtain apparent B vectors by rotating the observed E vector by $90^{\circ}$,
uncorrected for Faraday rotation.

For NGC~4388, NGC~4402, NGC~4438, and NGC~4501 our 20~cm total power
data have artifacts most likely due to the tenuous UV coverage. We
therefore preferred the VIVA H{\sc i} 20~cm continuum images (Chung
et al. 2009; for the more sophisticated data reduction of
NGC~4438 see Vollmer et al. 2009).

\begin{table*}
      \caption{Integration times and rms.}
         \label{tab:table}
      \[
         \begin{array}{llccccccccc}
           \hline
           \noalign{\smallskip}
           {\rm galaxy\ name } & {\rm m_{\rm B}^{(1)}} & i^{(2)} & {\rm Dist.^{(3)}} &{\rm integration}&{\rm integration} & {\rm bandwidth} & {\rm rms\ TP6cm} & {\rm rms\ TP20cm} & {\rm rms\ PI6cm} & {\rm rms\ PI20cm}\\
        &  & & & {\rm time\ (6cm)}& {\rm time\ (20cm)} & {\rm (20cm)} & (\mu{\rm Jy/} &(\mu{\rm Jy/} & (\mu{\rm Jy/} &(\mu{\rm Jy/}\\
        & {\rm (mag)} & {\rm (deg)}& {\rm (deg)} & {\rm (h:min)} & {\rm (h:min)}& {\rm (MHz)} & {\rm beam)} & {\rm beam)} & {\rm beam)} & {\rm beam)}\\
       \noalign{\smallskip}
       \hline
       \noalign{\smallskip}
       {\bf NGC~4321} & 10.02 & 27^{\rm (a)} & 4.0 & 7:45 & 0:30 & 2\times50 & 10 & 110 & 9 & 35\\
           \noalign{\smallskip}
       \hline
       \noalign{\smallskip}
       {\bf NGC~4388} & 11.87 & 77^{\rm (a)} & 1.3 & 9:25 & 8:00 & {\rm VIVA} & 60 & 300 & 9 & 200 \\
       \noalign{\smallskip}
           \hline
       \noalign{\smallskip}
       {\bf NGC~4396} & 13.07 & 72^{\rm (b)} & 3.5 & 8:00 & 0:30 & 2\times50 & 8 & 60 & 9 & 25 \\
       \noalign{\smallskip}
           \hline
       \noalign{\smallskip}
       {\bf NGC~4402} & 12.64 & 74^{\rm (c)} & 1.4 & 5:00 & 8:00 & {\rm VIVA} & 21 & 80 & 13 & 100 \\
       \noalign{\smallskip}
           \hline
       \noalign{\smallskip}
       {\bf NGC~4438} & 11.12 & 68^{\rm (a)}/85^{\rm (c)} & 1.0 & 7:45 & 8:00 &{\rm VIVA} & 30 & 60 & 14 & 300 \\
       \noalign{\smallskip}
           \hline
       \noalign{\smallskip}
       {\bf NGC~4501} & 10.50 & 57^{\rm (a)} & 2.0 & 3:55 & 5:00 &{\rm VIVA} & 14 & 220 & 11 &  50 \\
       \noalign{\smallskip}
           \hline
       \noalign{\smallskip}
       {\bf NGC~4535} & 10.73 & 43^{\rm (a)} & 4.3 & 9:00 & 0:30 & 2\times50 & 9 & 80 & 9 & 35 \\
       \noalign{\smallskip}
           \hline
       \noalign{\smallskip}
       {\bf NGC~4654} & 11.31 & 51^{\rm (a)} & 3.4 & 7:50 & 0:30 & 2\times50 & 9 & 50 & 9 & 30 \\
       \noalign{\smallskip}
           \hline
        \end{array}
      \]
\begin{list}{}{}
\item[$^{(1)}$ This research has made use of the GOLD Mine]
\item[\ \ \ \ Database (Gavazzi et al. 2003)]
\item[$^{(2)}$ Inclination angle; $^{(3)}$ Distance from M~87]
\item[$^{(a)}$ from Cayatte et al. (1990); $^{(b)}$ from NED]
\item[$^{(c)}$ from Kenney et al. (1995)]
\end{list}
\end{table*}

\section{Results\label{sec:results}}

\begin{table}
      \caption{Integrated flux densities.}
         \label{tab:tableflux}
      \[
         \begin{array}{lrrrr}
           \hline
           \noalign{\smallskip}
           {\rm galaxy\ name } & S_{\rm TP20cm} & S_{\rm PI20cm} & S_{\rm TP6cm} & S_{\rm PI6cm} \\
	    & {\rm mJy} & {\rm mJy} & {\rm mJy} & {\rm mJy} \\
       \noalign{\smallskip}
       \hline
       \noalign{\smallskip}
       {\bf NGC~4321} & 256 \pm 2 & 17.2 \pm 0.5 & 99.1 \pm 0.2 & 14.1 \pm 0.2 \\
       \noalign{\smallskip}
       \hline
       \noalign{\smallskip}
       {\bf NGC~4388} & 178 \pm 2 & - & 81.6 \pm 0.5 & 2.3 \pm 0.1 \\
       \noalign{\smallskip}
       \hline
       \noalign{\smallskip}
       {\bf NGC~4396} & 15.8 \pm 0.5 & 1.5 \pm 0.2 & 8.3 \pm 0.1 & 1.0 \pm 0.1 \\
       \noalign{\smallskip}
       \hline
       \noalign{\smallskip}
       {\bf NGC~4402} & 68 \pm 2 & - &  26.3 \pm 0.2 & 2.5 \pm 0.1 \\
       \noalign{\smallskip}
       \hline
       \noalign{\smallskip}
       {\bf NGC~4438} & 102 \pm 3 & - & 73.9 \pm 0.3 & 6.3 \pm 0.1 \\
       \noalign{\smallskip}
       \hline
       \noalign{\smallskip}
       {\bf NGC~4501} & 331 \pm 2 & 10.4 \pm 0.4 & 96.2 \pm 0.2 & 11.2 \pm 0.1 \\
       \noalign{\smallskip}
       \hline
       \noalign{\smallskip}
       {\bf NGC~4535} & 66 \pm  1 & 13.9 \pm 0.4 & 28.7 \pm 0.2 & 4.2 \pm 0.1  \\
       \noalign{\smallskip}
       \hline
       \noalign{\smallskip}
       {\bf NGC~4654} & 134.6 \pm 0.5 & 6.0 \pm 0.3 & 49.9 \pm 0.1 & 3.1 \pm 0.1 \\
       \noalign{\smallskip}
       \hline
         \end{array}
	 \]
\end{table}

\subsection{Galaxy properties \label{sec:properties}}

\subsubsection{NGC~4321}

\begin{figure*}
  \centering
  \includegraphics[width=16cm]{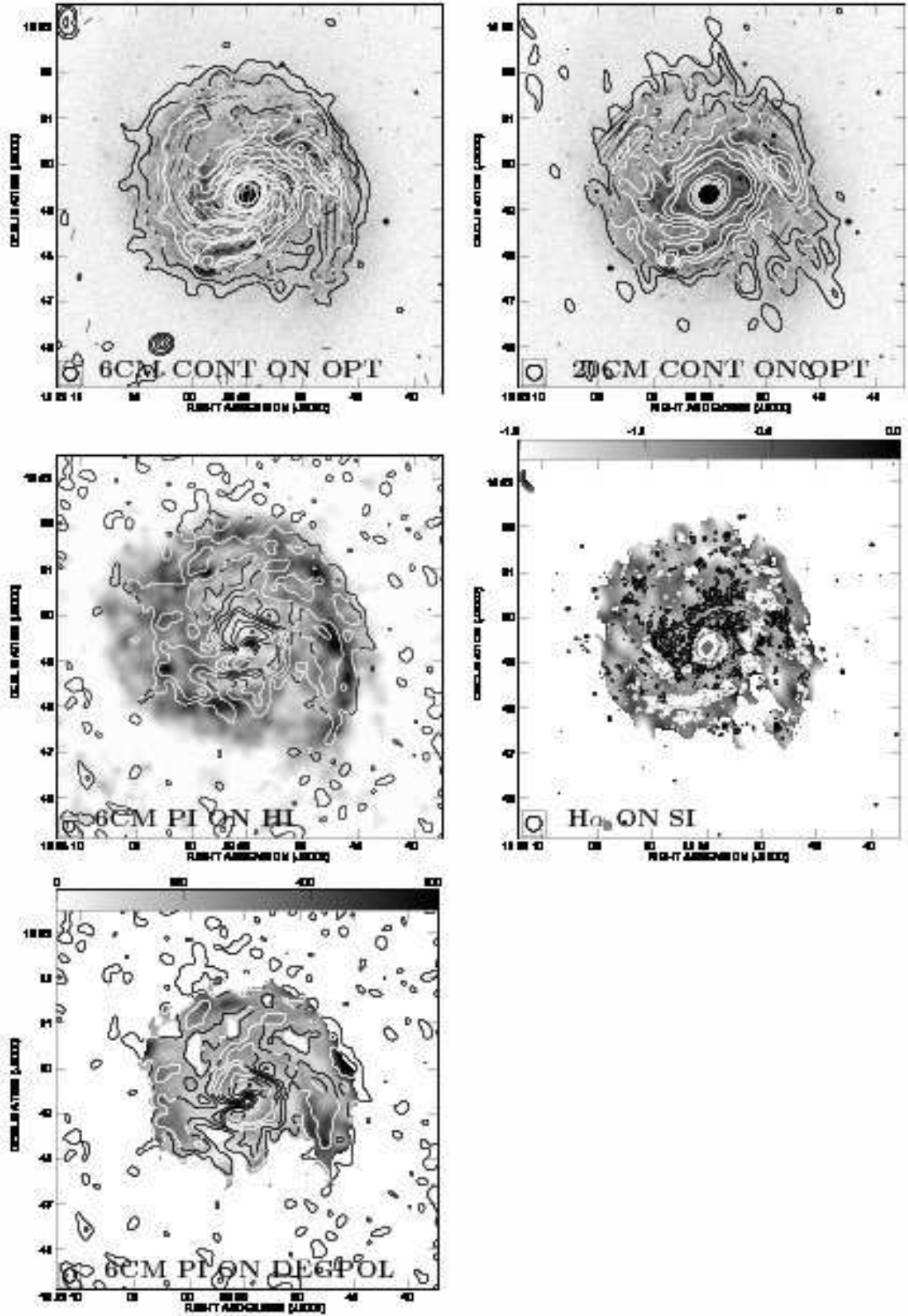}
  \caption{NGC~4321: 
    From top left to bottom right: 6~cm total power emission distribution on DSS B band image
    together with the apparent B vectors,
    20~cm total power emission distribution on DSS B band image together with the apparent B vectors,
    6~cm polarized emission distribution on VIVA H{\sc i} distribution,
    H$\alpha$ emission distribution on spectral index map, and 
    6~cm polarized emission distribution on degree of polarization.
    6cm continuum contour levels are 70~$\mu$Jy $\times 
    (1,2,4,6,8,10,20,30,40,50)$. 20cm continuum contour levels are
    330~$\mu$Jy $\times (1,2,4,6,8,10,20,30,40,50)$.
    6~cm polarized intensity contour levels are 10~$\mu$Jy $\times
    (4,8,12,16,20,30,40,50)$. 
    The H$\alpha$ contours on the spectral index map appear white if
    the spectral index $\alpha > -0.8$ where $S \propto \nu^{\alpha}$.
    The indicated levels of the degree of polarization are in units of 0.1\,\%.
   }
  \label{fig:n4321}
\end{figure*}

NGC~4321 is the only galaxy in our sample which shows relatively symmetric emission
distributions at all wavelengths as it is observed in field spiral galaxies (Fig.~\ref{fig:n4321}).
The 6~cm and 20~cm total power distributions follow the stellar and gas distribution.
The radio continuum spectrum is flat ($\alpha > -0.8$ with $S_{\nu} \propto
\nu^{\alpha}$) in the regions of the main spiral arms and steep ($\alpha \leq -0.8$)
elsewhere.
The polarized emission and the degree of polarization are highest in the
interarm regions.
In addition, the distribution of  the degree of polarization shows an azimuthally symmetric 
radial gradient with a higher degree of polarization towards the outer disk. 
All characteristics cited above are typical for an unperturbed spiral galaxy (Beck 2005
and references therein).
The regions of highest degrees of polarization (up to 40\%) are found
at the western edge of the polarized emission distribution.

\subsubsection{NGC~4388}

\begin{figure*}
  \centering
  \includegraphics[width=16cm]{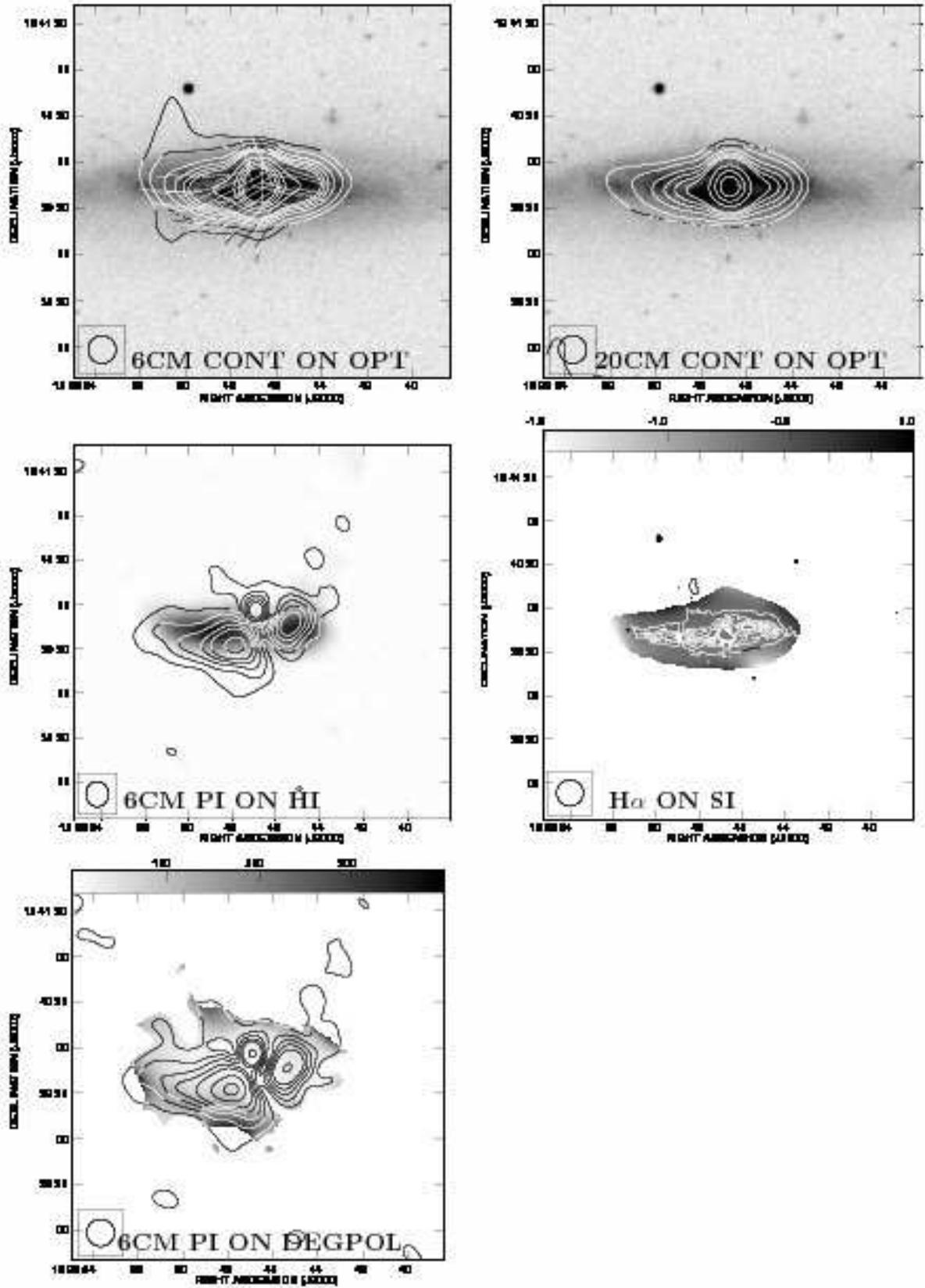}
  \caption{NGC~4388:
    From top left to bottom right: 6~cm total power emission distribution on DSS B band image
    together with the apparent B vectors,
    20~cm total power emission distribution on DSS B band image together with the apparent B vectors,
    6~cm polarized emission distribution on VIVA H{\sc i} distribution,
    H$\alpha$ emission distribution on spectral index map, and 
    6~cm polarized emission distribution on degree of polarization.
    6cm continuum contour levels are 320~$\mu$Jy $\times 
    (1,2,4,6,8,10,20,30,40,50)$. 20cm continuum contour levels are
    1500~$\mu$Jy $\times (1,2,4,6,8,10,20,30,40,50)$.
    6~cm polarized intensity contour levels are 10~$\mu$Jy $\times
    (4,8,12,16,20,30,40,50)$.
    The H$\alpha$ contours on the spectral index map appear white if
    the spectral index $\alpha > -0.8$ where $S \propto \nu^{\alpha}$.
    The indicated levels of the degree of polarization are in units of 0.1\,\%.
   }
  \label{fig:n4388}
\end{figure*}

NGC~4388 harbors a Seyfert~2 nucleus with an associated bright radio source 
and a $15''$-long jet to the north (Hummel \& Saikia 1991).
The disk and jet emission perpendicular to the galactic plane has already been detected at 
6cm by Hummel et al. (1983).
In Fig.~\ref{fig:n4388} we observe emission from the nuclear outflow in polarized 6~cm emission with vertical 
magnetic field vectors north of the nucleus.
The radio continuum disk emission at 20~cm and 6~cm and the H{\sc i} emission
distribution are truncated at about half the optical radius. 
These emission distributions are more extended to the east
than to the west. The 6~cm total power extension perpendicular to the
major axis east of the nucleus needs confirmation. 
It is present in observations made on two separate days.
We do not detect a disk-wide radio halo.
The eastern 6~cm polarized emission maximum is located $\sim 20''$
south of the disk plane, on the edge of the H{\sc i} emission.
The western 6~cm polarized emission maximum is slightly elongated
parallel to the minor axis.
The radio continuum spectrum is flat ($\alpha > -0.8$)
everywhere in the disk. We observe an asymmetric distribution of the degree
of polarization at 6~cm with 10\,\% of polarization at the southeastern edge, 
whereas the degree of polarization is
much less at the northern edge of the galactic disk.

\subsubsection{NGC~4396}

\begin{figure*}
  \centering
  \includegraphics[width=16cm]{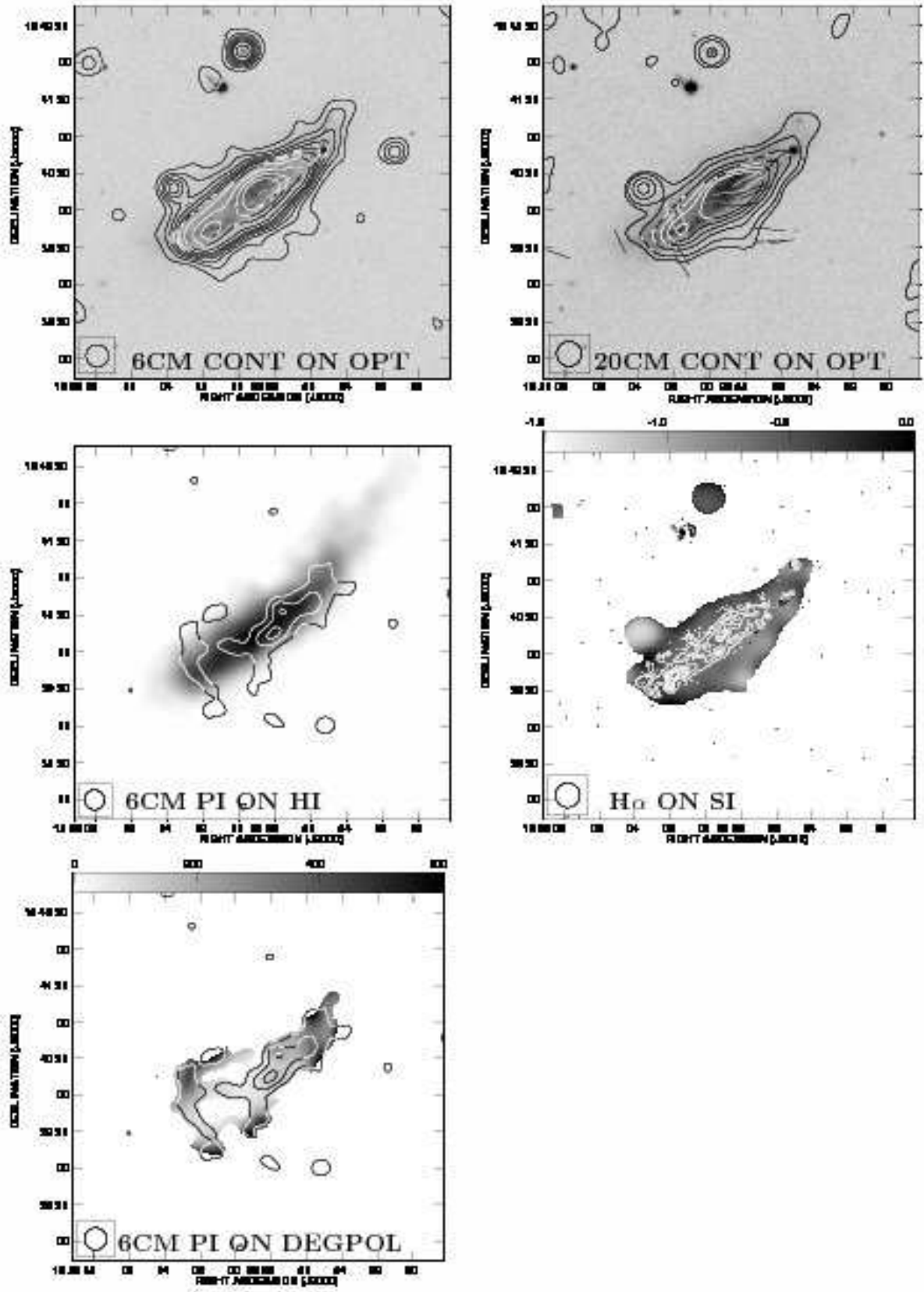}
  \caption{NGC~4396: 
    From top left to bottom right: 6~cm total power emission distribution on DSS B band image
    together with the apparent B vectors,
    20~cm total power emission distribution on DSS B band image together with the apparent B vectors,
    6~cm polarized emission distribution on VIVA H{\sc i} distribution,
    H$\alpha$ emission distribution on spectral index map, and 
    6~cm polarized emission distribution on degree of polarization.
    6cm continuum contour levels are 40~$\mu$Jy $\times 
    (1,2,3,4,5,6,8,12,16,20,30,40,50)$. 20cm continuum contour levels are
    150~$\mu$Jy $\times (1,2,4,6,8,10,20,30,40,50)$.
    6~cm polarized intensity contour levels are 8~$\mu$Jy $\times
    (4,8,12,16,20,30,40,50)$. 
    The H$\alpha$ contours on the spectral index map appear white if
    the spectral index $\alpha > -0.8$ where $S \propto \nu^{\alpha}$.
    The indicated levels of the degree of polarization are in units of 0.1\,\%.
   }
  \label{fig:n4396}
\end{figure*}

The 20~cm and 6~cm total power emission of NGC~4396 is truncated within the 
optical disk and shows a tail structure to the northwest (Fig.~\ref{fig:n4396}).
As the H{\sc i} tail, the radio continuum tail is bend to the north, but it is less 
extended than the H{\sc i} tail. We do not detect a disk-wide radio halo.
The 6~cm polarized emission is mainly found
in the southeast of the galactic disk extending into the tail.
Faint polarized emission perpendicular to the galactic disk is
found in the eastern region. The radio continuum spectrum is flat over the whole starforming
disk of NGC~4396. 
The distribution of the degree of polarization at 6~cm increases towards the tail
with a maximum value of 30\,\% at the northwestern tip.
It also increases to the north in the eastern polarized emission feature.

\subsubsection{NGC~4402}

\begin{figure*}
  \centering
  \includegraphics[width=16cm]{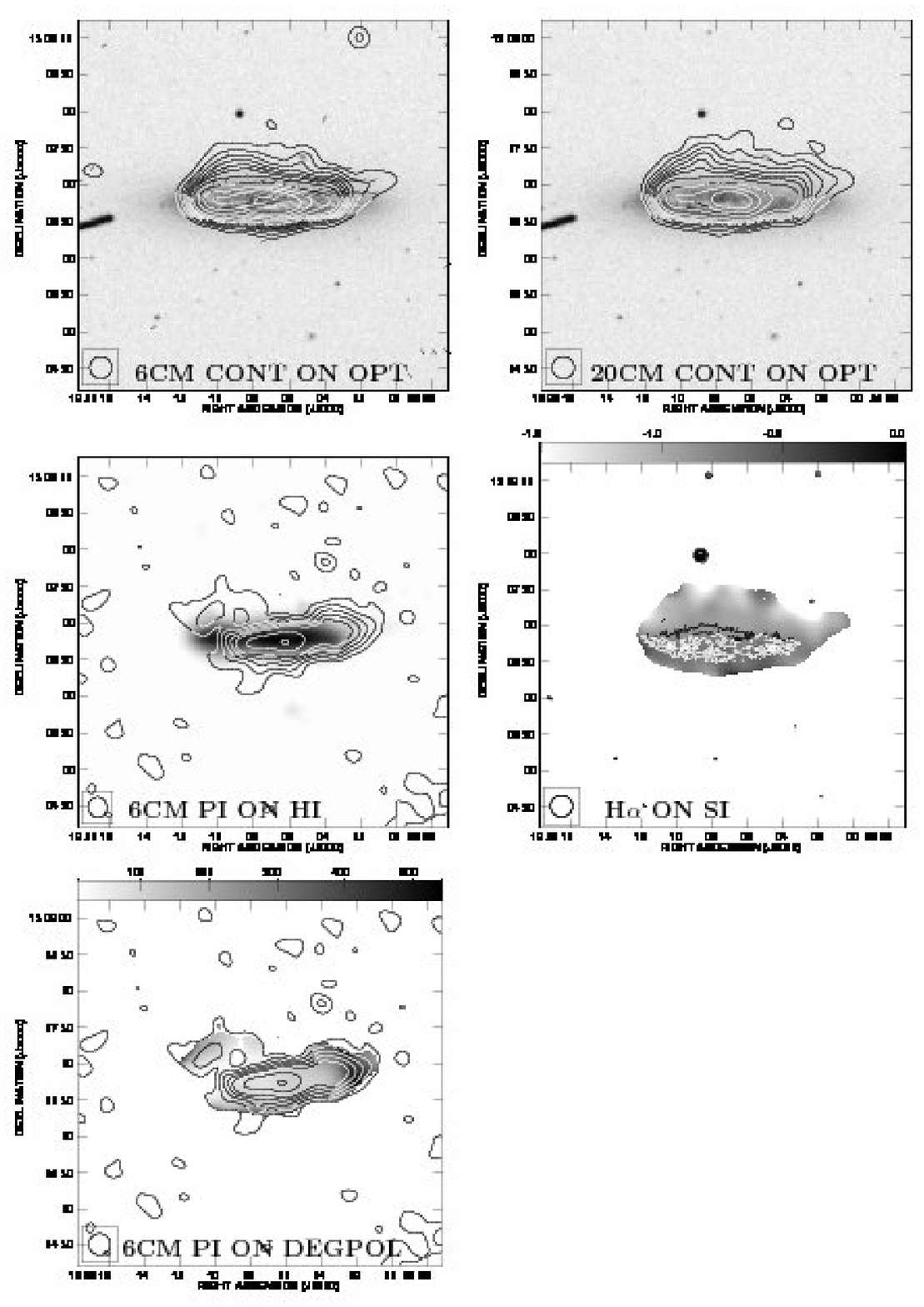}
  \caption{NGC~4402: 
    From top left to bottom right: 6~cm total power emission distribution on DSS B band image
    together with the apparent B vectors,
    20~cm total power emission distribution on DSS B band image together with the apparent B vectors,
    6~cm polarized emission distribution on VIVA H{\sc i} distribution,
    H$\alpha$ emission distribution on spectral index map, and 
    6~cm polarized emission distribution on degree of polarization.
    6cm continuum contour levels are 120~$\mu$Jy $\times 
    (1,2,3,4,5,6,8,12,16,20,30,40,50)$. 20cm continuum contour levels are
    400~$\mu$Jy $\times (1,2,3,4,5,6,8,12,16,20,30,40,50)$.
    6~cm polarized intensity contour levels are 8~$\mu$Jy $\times
    (4,8,12,16,20,30,40,50)$. 
    The H$\alpha$ contours on the spectral index map appear white if
    the spectral index $\alpha > -0.8$ where $S \propto \nu^{\alpha}$.
    The indicated levels of the degree of polarization are in units of 0.1\,\%.
   }
  \label{fig:n4402}
\end{figure*}

As the H{\sc i} emission distribution, the 20~cm and 6~cm total
power emission distribution are truncated at about half of the optical radius 
(Chung et al. 2009, Fig.~\ref{fig:n4402}).
In addition, we observe a disk-wide extension to the north,
whereas the southern edge of the emission distributions is sharp
(see also Crowl et al. 2005).
Although the total power disk emission is symmetric along the major axis, the 6~cm polarized
emission is more extended to the west than to the east.
Moreover, we find faint extraplanar polarized emission northeast of the galactic disk.
The extended extraplanar radio continuum emission north of the
disk shows a steepening of the radio continuum spectrum.
The distribution of the degree of polarization has a north--south and an east--west gradient, the
latter being dominant. At the southern edge the degree of
polarization is 10\,\%, whereas it increases up to 40\,\% at the
western edge of the galactic disk.

\subsubsection{NGC~4438}

\begin{figure*}
  \centering
  \includegraphics[width=16cm]{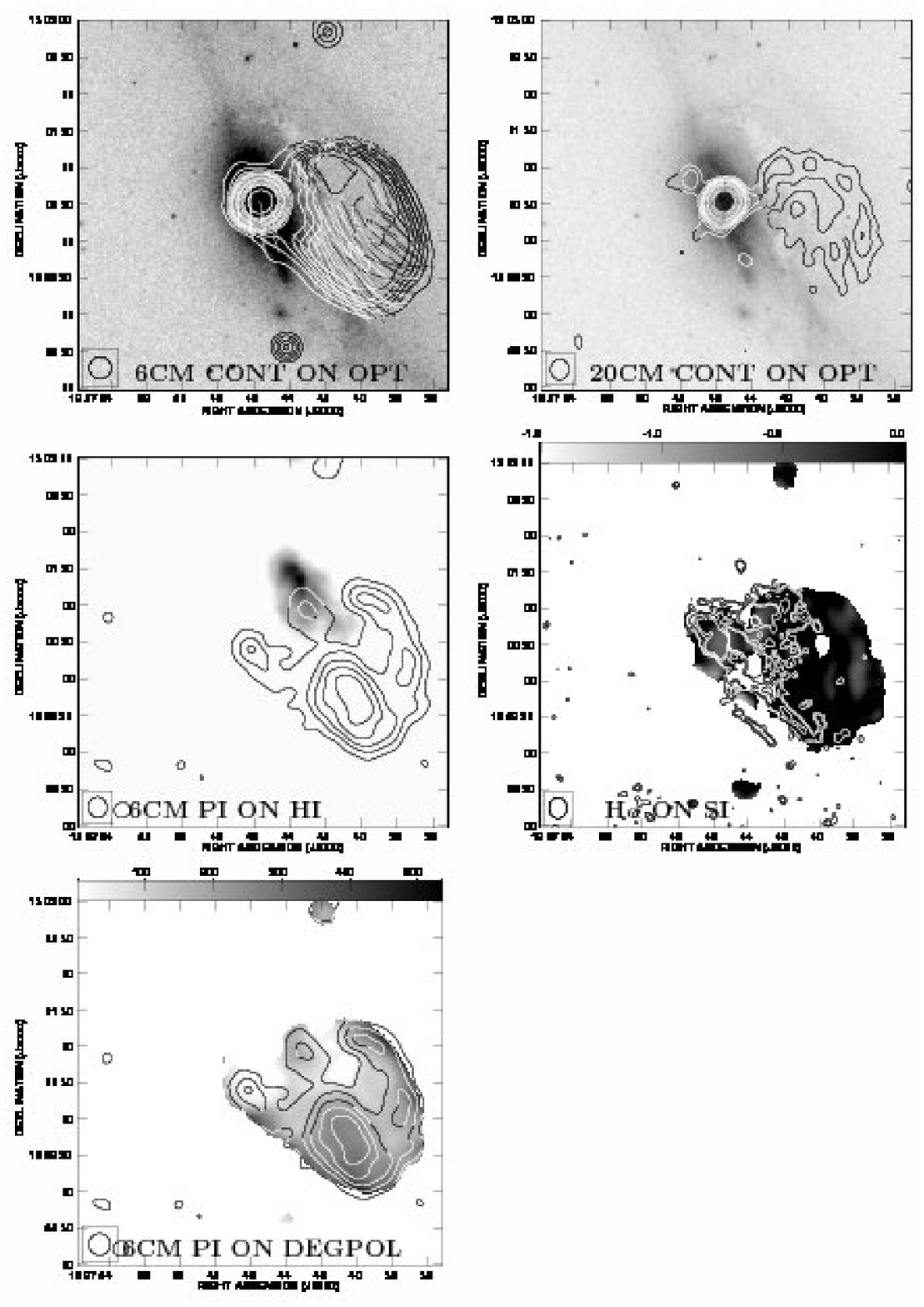}
  \caption{NGC~4438: 
    From top left to bottom right: 6~cm total power emission distribution on DSS B band image
    together with the apparent B vectors,
    20~cm total power emission distribution on DSS B band image together with the apparent B vectors,
    6~cm polarized emission distribution on VIVA H{\sc i} distribution,
    H$\alpha$ emission distribution on spectral index map, and 
    6~cm polarized emission distribution on degree of polarization.
    6cm continuum contour levels are 160~$\mu$Jy $\times 
    (1,2,3,4,5,6,8,12,16,20,30,40,50)$. 20cm continuum contour levels are
    180~$\mu$Jy $\times (1,2,3,4,5,6,8,12,16,20,30,40,50)$.
    6~cm polarized intensity contour levels are 15~$\mu$Jy $\times
    (4,8,12,16,20,30,40,50)$. 
    The H$\alpha$ contours on the spectral index map appear white if
    the spectral index $\alpha > -0.8$ where $S \propto \nu^{\alpha}$.
    The indicated levels of the degree of polarization are in units of 0.1\,\%.
   }
  \label{fig:n4438}
\end{figure*}

NGC~4438 harbors a Seyfert 2 nucleus with an associated strong nuclear radio source
(Hummel \& Saikia 1991).
In addition, we observe prominent extraplanar total power emission at
20~cm and 6~cm extending further than emission at other wavelengths (Fig.~\ref{fig:n4438}).
The 6~cm polarized extraplanar emission has a shell-like distribution with a
pronounced maximum in the south. We also observe polarized emission from 
the galactic disk south from the nucleus.
The spectral index of the western extraplanar emission region of
NGC~4438 is constant. Its value is uncertain because of M~87's
strong sidelobes at 20cm which makes the 20~cm flux density uncertain. 
In Vollmer et al. (2009) we argue that we miss a substantial part of the flux
density of NGC~4438 at 20~cm due to these sidelobes and that
the spectral index is most likely $\alpha = -0.8$, i.e. the typical
value for synchrotron emission. 
There is a positive gradient of the degree of
polarization towards the border of the extraplanar synchrotron
emission where the degree of polarization increases to $\sim
30$\,\%.  The southern half of the extraplanar
radio emission shows a degree of polarization of 20\,\%.

\subsubsection{NGC~4501}

\begin{figure*}
  \centering
  \includegraphics[width=16cm]{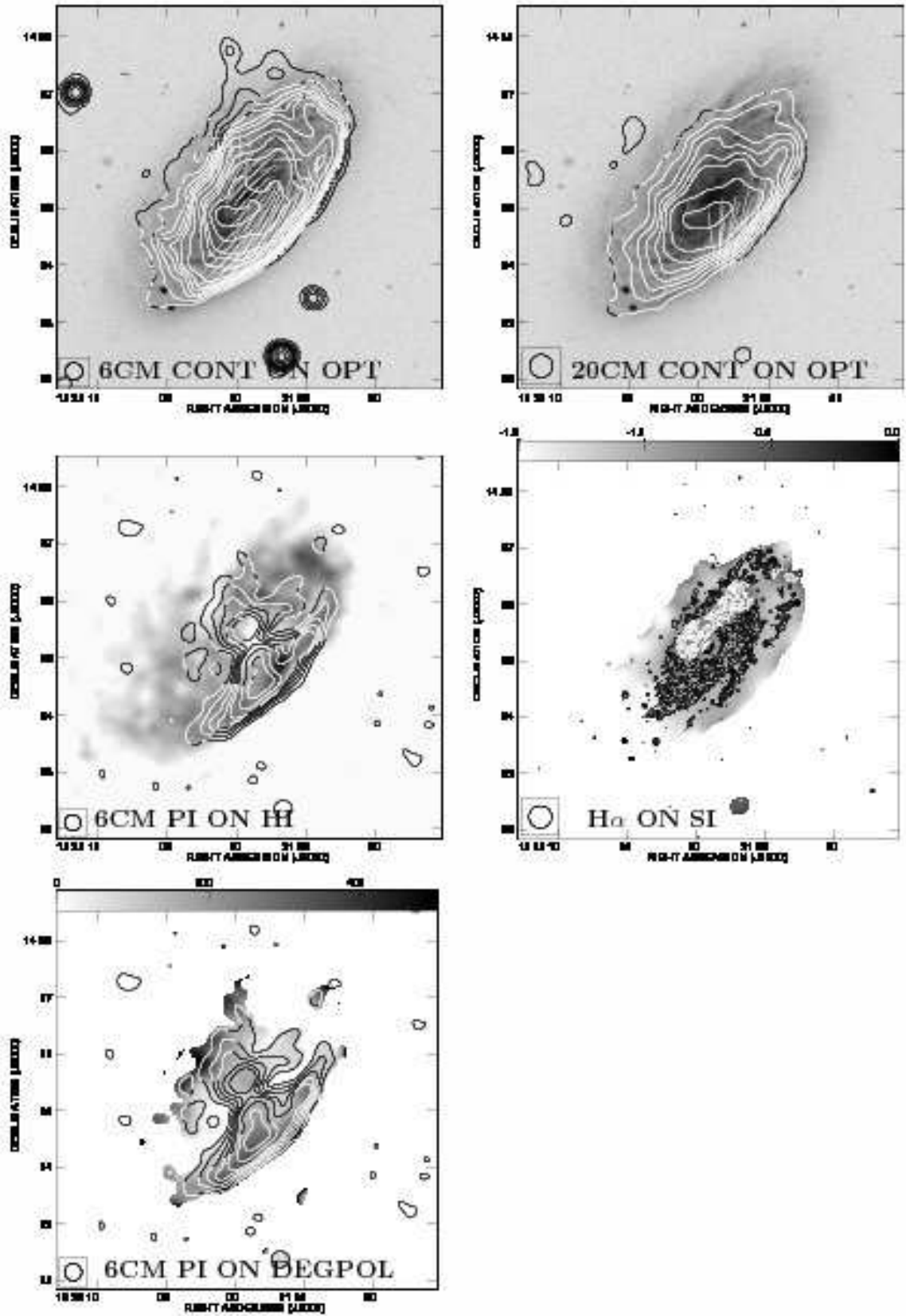}
  \caption{NGC~4501: 
    From top left to bottom right: 6~cm total power emission distribution on DSS B band image
    together with the apparent B vectors,
    20~cm total power emission distribution on DSS B band image together with the apparent B vectors,
    6~cm polarized emission distribution on VIVA H{\sc i} distribution,
    H$\alpha$ emission distribution on spectral index map, and 
    6~cm polarized emission distribution on degree of polarization.
    6cm continuum contour levels are 80~$\mu$Jy $\times 
    (1,2,3,4,5,6,8,12,16,20,30,40,50)$. 20cm continuum contour levels are
    1500~$\mu$Jy $\times (1,2,3,4,5,6,8,12,16,20,30,40,50)$.
    6~cm polarized intensity contour levels are 15~$\mu$Jy $\times
    (4,8,12,16,20,30,40,50)$. 
    The H$\alpha$ contours on the spectral index map appear white if
    the spectral index $\alpha > -0.8$ where $S \propto \nu^{\alpha}$.
    The indicated levels of the degree of polarization are in units of 0.1\,\%.
   }
  \label{fig:n4501}
\end{figure*}

As the H{\sc i} distribution, the 20~cm and 6~cm total power emission distributions are 
truncated at about the optical radius (Fig.~\ref{fig:n4501}).
Whereas the southwestern edge is sharp, there is extended emission to the 
northeast. The 6~cm polarized emission shows a long ridge extending over
almost the entire south western edge of the galactic disk.
There is a secondary maximum north of the nucleus.
Whereas the radio continuum spectrum is flat ($\alpha > -0.8$) in the northeastern spiral arm,
it is steep ($\alpha \leq -0.8$) in the southwestern spiral arms.
We observe a general steepening of the spectrum to the southeast
of the galactic disk.
The degree of polarization shows an asymmetric distribution. It increases towards the
southwestern and northeastern edges of the galactic disk. Whereas it
rises to 20\,\% in the northeast, it increases to 30\,\% in the southwest.

\subsubsection{NGC~4535}

\begin{figure*}
  \centering
  \includegraphics[width=16cm]{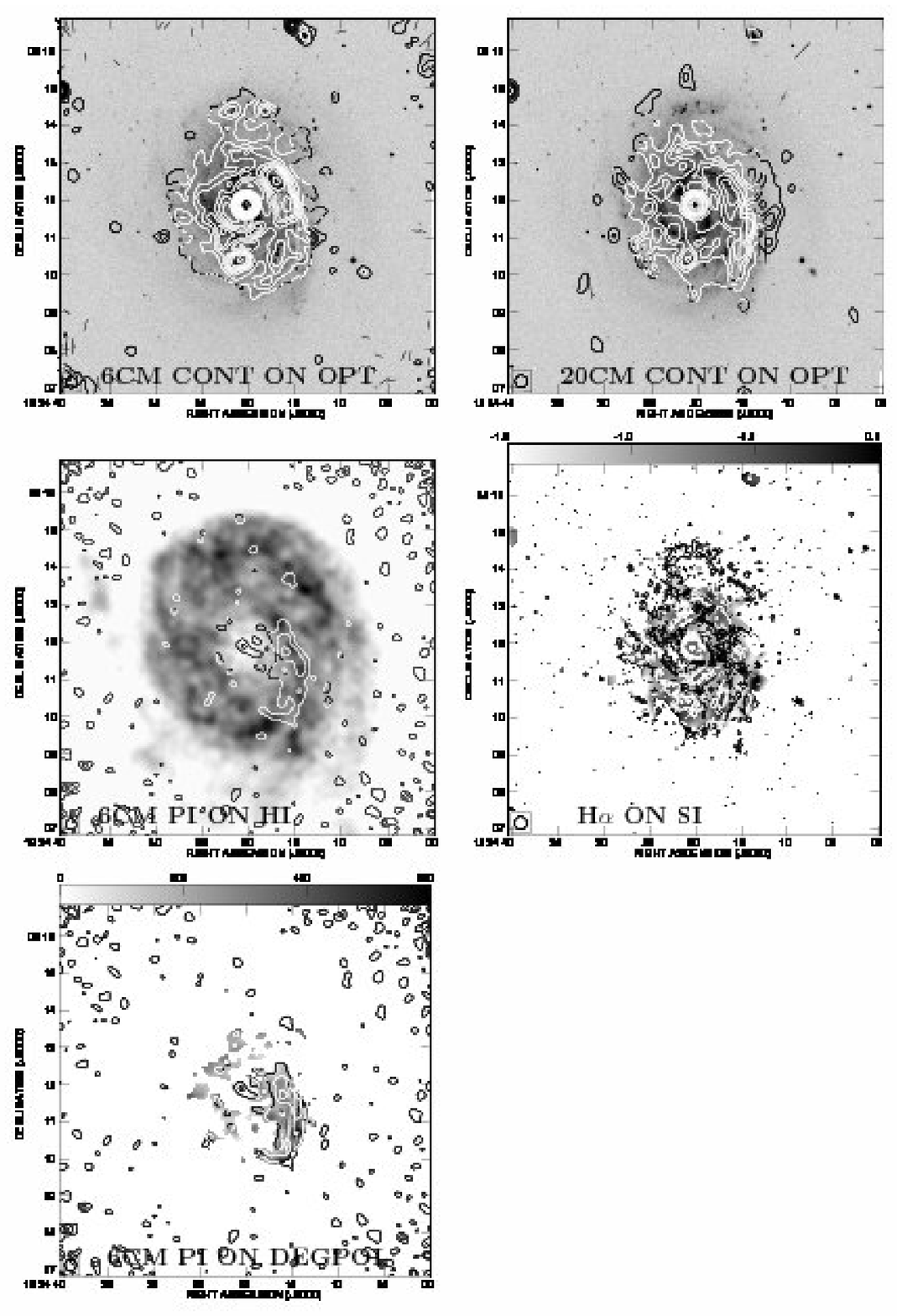}
  \caption{NGC~4535:
    From top left to bottom right: 6~cm total power emission distribution on DSS B band image
    together with the apparent B vectors,
    20~cm total power emission distribution on DSS B band image together with the apparent B vectors,
    6~cm polarized emission distribution on VIVA H{\sc i} distribution,
    H$\alpha$ emission distribution on spectral index map, and 
    6~cm polarized emission distribution on degree of polarization.
    6cm continuum contour levels are 55~$\mu$Jy $\times 
    (1,2,3,4,5,6,8,12,16,20,30,40,50)$. 20cm continuum contour levels are
    220~$\mu$Jy $\times (1,2,3,4,6,8,12,16,20,30,40,50)$.
    6~cm polarized intensity contour levels are 10~$\mu$Jy $\times
    (4,8,12,16,20,30,40,50)$. 
    The H$\alpha$ contours on the spectral index map appear white if
    the spectral index $\alpha > -0.8$ where $S \propto \nu^{\alpha}$.
    The indicated levels of the degree of polarization are in units of 0.1\,\%.
   }
  \label{fig:n4535}
\end{figure*}

The 20~cm and 6~cm total power emission extends to $\sim 0.7$ times
the optical radius, well inside the H{\sc i} distribution (Fig.~\ref{fig:n4535}). 
The total power emission associated with the western optical arm is stronger than that 
of the rest of the disk.
Polarized 6~cm continuum emission is only detected in the region of the 
western optical arm. 
The radio continuum spectrum is flat ($\alpha > -0.8$) in the regions of the starforming
spiral arms and steep  ($\alpha \leq -0.8$) elsewhere.
A high degree of polarization ($\sim 40$\,\%) is found in the southern
part of the polarized emission arm.

\subsubsection{NGC~4654}

\begin{figure*}
  \centering
  \includegraphics[width=16cm]{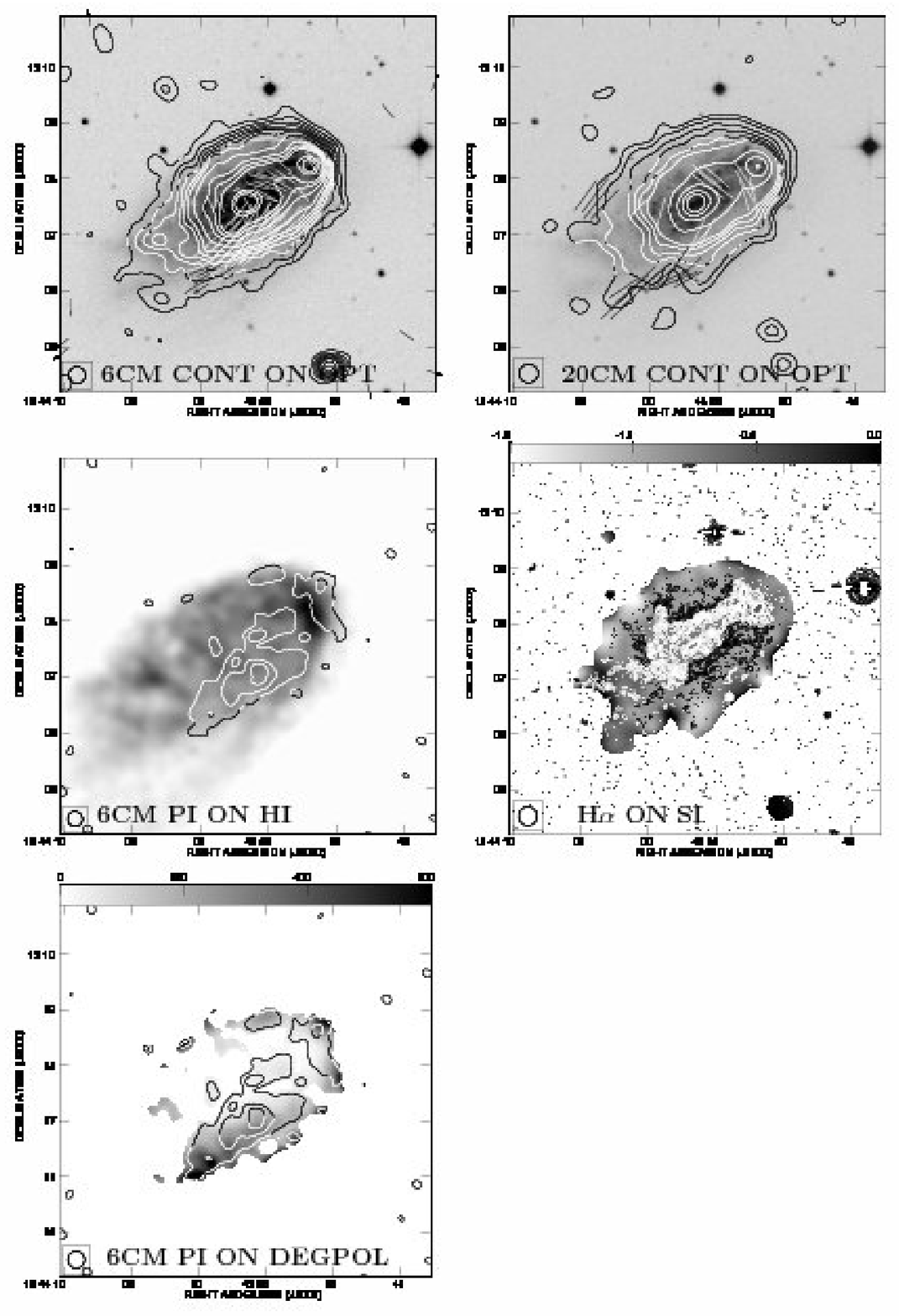}
  \caption{NGC~4654: 
    From top left to bottom right: 6~cm total power emission distribution on DSS B band image
    together with the apparent B vectors,
    20~cm total power emission distribution on DSS B band image together with the apparent B vectors,
    6~cm polarized emission distribution on VIVA H{\sc i} distribution,
    H$\alpha$ emission distribution on spectral index map, and 
    6~cm polarized emission distribution on degree of polarization.
    6cm continuum contour levels are 57~$\mu$Jy $\times 
    (1,2,4,6,8,10,20,30,40,50)$. 20cm continuum contour levels are
    180~$\mu$Jy $\times (1,2,3,4,6,8,12,16,20,30,40,50)$.
    6~cm polarized intensity contour levels are 10~$\mu$Jy $\times
    (4,8,12,16,20,30,40,50)$. Optical DSS and VIVA H{\sc i} images are used.
    The H$\alpha$ contours on the spectral index map appear white if
    the spectral index $\alpha > -0.8$ where $S \propto \nu^{\alpha}$.
    The indicated levels of the degree of polarization are in units of 0.1\,\%.
   }
  \label{fig:n4654}
\end{figure*}

The 20~cm and 6~cm total power emission roughly follows the stellar
distribution of the galactic disk (Fig.~\ref{fig:n4654}). Along the major axis, the emission
is stronger to the northwest than to the southeast.
We observe a sharp edge of the emission distribution to the
northwest and some faint extended emission in the direction of the prominent
southeastern H{\sc i} tail. The 6~cm polarized emission distribution is
asymmetric with a maximum south of the galaxy center well within the
H{\sc i} distribution. 
The radio continuum spectrum is flat ($\alpha > -0.8$) in the regions of the 
starforming spiral arms and steep ($\alpha \leq -0.8$) elsewhere.
The degree of 6~cm polarization increase to $\sim 20$\,\% towards
the southern edge of the polarized emission distribution.
The highest degrees of polarization ($\sim 30$\,\%) are found at the south
eastern tip.

\section{Comparison between galaxies\label{sec:comparison}}

In this section we compare the total power distributions at 20/6~cm
to the gas and stellar disks followed by the polarized radio continuum
emission and the spectral index. 
The degree of polarization at 6~cm and the rotation
measure between 20 and 6~cm are presented at the end of this
section.

\subsection{Total power emission\label{sec:totalpower}}

The total power emission is a mixture of synchrotron and thermal emission.
The intensity of the synchrotron emission depends on the density of
cosmic-ray electrons and the square of the total magnetic field in the sky plane.
The density of thermal and of cosmic ray electrons depends on the star formation activity of the galaxy.
In an isolated spiral galaxy the total magnetic field is
dominated by the turbulent small-scale component.
As the result of ram pressure compression the total power emission can be locally enhanced by
(i) an increased local star formation rate and/or (ii) compression of the
magnetic field.

Only NGC~4321 shows a symmetric total power distribution as it is
observed in unperturbed field spiral galaxies.
The 20cm and 6cm total power distributions of two (NGC~4501, NGC~4654) 
out of three mildly inclined spiral galaxies show sharp edges with steep gradients at
one side of the disk. These edges coincide with H{\sc i} edges and
are due to ram pressure (Soida et al. 2006, Vollmer et al. 2008).
Murphy et al. (2009) combined Spitzer FIR and VLA 20cm radio continuum imaging
to study the FIR-radio correlation in Virgo spiral galaxies. 
For 6 out of 10 sample galaxies, they found regions along their outer edges that are highly deficient 
in the radio compared with models relying on the FIR-radio correlation of field galaxies.
The observed sharp edges in the total power distribution might be linked to this phenomenon 
of radio deficient edges.
\\
We observe asymmetric extraplanar emission in 3 out of 4 edge-on
galaxies (NGC~4396, NGC~4402, NGC~4438; see Sect.~\ref{sec:dwh}). Whereas the extraplanar
radio continuum emission follows the H{\sc i} emission in the northwest of
NGC~4396, it is more extended than the H{\sc i} emission in NGC~4402
and NGC~4438. At our limiting flux density level there is no total power 
emission associated with the
ionized nuclear outflow in NGC~4388 (Veilleux et al. 1999). In all
moderately inclined galaxies (NGC~4321, NGC~4501, NGC~4535, NGC~4654)
the 20~cm and 6~cm emission follow the recent massive star formation as
observed in the H$\alpha$ line. The
southeastern H{\sc i} tail of NGC~4654 (Phookun \& Mundy 1995) has
no radio continuum counterpart. This is most probably due to the
lack of star formation in this tail.

\subsection{Polarized continuum and H{\sc i} emission
\label{sec:polHI}}

Polarized continuum emission is caused by the regular large-scale
magnetic field. Polarized emission can be enhanced by large-scale shear or compression motions.

Only NGC~4321 has a
relatively symmetric polarized emission distribution which is
highest in the interarm regions. All other galaxies have asymmetric elongated
ridges of polarized emission located in the outer parts of the
galactic disks (Vollmer et al. 2007). These ridges are located within the H{\sc i}
distribution, except for NGC~4438, where the western local maximum of
polarized emission is located within the molecular gas disk (Vollmer
et al. 2005). Most of the ridges are close to the outer edge of the H{\sc
i} distribution. Especially in NGC~4501 the H{\sc i} and polarized
emission distributions show a sharp edge to the southeast where ram
pressure acts on the galaxy (see Vollmer et al. 2008). Such a sharp
edge is also observed in the northeast of NGC~4654 (see Soida et
al. 2006). Only the southern polarized ridge of NGC~4654 and that of
NGC~4535 are well inside the H{\sc i} distribution. Given the
asymmetric velocity field of NGC~4535 in the region of the maximum of
polarized radio continuum emission (Chung et al. 2009),
the enhancement of the polarized emission is most probably due to shear motions.
Moreover, all edge-on galaxies show extraplanar polarized emission which extends
further than the H{\sc i} emission. In NGC~4388 this extraplanar polarized
emission is probably due to the AGN outflow (Veilleux et al. 1999).
In NGC~4396 it is located on the southeast and extends to the north.
In NGC~4402 it is located in the western side of the disk where the action of ram
pressure is maximum (Crowl et al. 2005). Whereas the extraplanar
polarized emission is faint in NGC~4396 and NGC~4402, it is
prominent in NGC~4438, extending further than emission at any other
wavelength.

The polarized emission at 20~cm suffers severe Faraday rotation and
depolarization, especially in edge-on galaxies. We will use this emission 
only to calculate the rotation measure in Sect.~\ref{sec:rm}.

\subsection{Spectral index\label{sec:spectralindex}}

As a general trend, we find flat radio continuum spectra
($\alpha > -0.8$) associated with regions of recent massive star
formation, indicating an enhanced fraction of thermal electrons. This corresponds to the
classical behavior of unperturbed spiral galaxies 
(e.g., Sukumar et al. 1987, Berkhuijsen et al. 2003, Tabatabaei et al. 2007a/b,
Heesen et al. 2009). For example, in NGC~4396 the radio continuum spectrum
of the northwestern extraplanar emission region is 
flat ($\alpha \sim -0.6$), because of its active star formation.
On the other hand, the extraplanar radio continuum emission north of
NGC~4402 shows a steepening of the spectrum due to the aging
of the relativistic electrons. This is consistent with the scenario
of Crowl et al. (2005) where it is assumed that the radio halo is compressed 
on the southern side and pushed out of the galactic disk on the
northern side. Surprisingly, the same steepening of the spectrum
is seen on the southwestern side of NGC~4501 where ram
pressure compresses the gas (Vollmer et al. 2008). This is contrary
to what we found in NGC~4522 (Vollmer et al. 2004) where we observed
a spectral flattening in the compressed region. 
Thus, in our sample we do not observe a flattening of the spectrum
in regions of enhanced polarized emission. A shock-induced reacceleration of
relativistic electrons as proposed by V\"{o}lk \& Xu (1994) is still a 
probable explanation for the flat spectrum associated to the
ridge of polarized emission in NGC~4522.

\subsection{Degree of polarization\label{sec:polarizedemission}}

The degree of polarization is defined as the ratio between polarized and
total power (mostly synchrotron) emission. The degree of polarization is a measure for the
fraction of the regularly oriented, large-scale magnetic field with respect to the
total magnetic field.

In our galaxy sample the degree of polarization varies between 10 and
40\,\%. In the face-on galaxy NGC~4321 the degree of polarization is highest in the
interarm regions as it is expected for an unperturbed galaxy.
Moreover, we observe an azimuthally symmetric radial gradient
as it is observed in field spirals (e.g., M~83: Neininger et al. 1993,
NGC~6946: Ehle \& Beck 1993). 
As an example for an edge-on spiral galaxy,
Dumke \& Krause (1998) combined 6~cm Effelsberg and VLA data of NGC~891 
(optical radius $R_{25}=13.5'$). 
They found a symmetrically increasing degree of polarization towards the
edges of the emission distribution to $\sim 10$\,\%.

In our sample, NGC~4388, NGC~4396, and NGC~4402 show vertically asymmetric
gradients of the degree of polarization. 
In addition, NGC~4396, NGC~4402, NGC~4501, and NGC~4654 show azimuthally
asymmetric distributions of the degree of polarization.
The rising degree of polarization towards the tails in NGC~4396 and NGC~4654
might be due to magnetic field ordering or shear motions in the gas
which is pulled away from or re-accreting onto the galactic disk.
Thus 5 out of 8 sample galaxies show these asymmetries.
This is different from the behavior of
field spirals (Beck 2005 and references therein, e.g. Neininger 1992, Sukumar \& Allen 1991) 
and is thus most probably due to the interaction between the galaxy and the cluster environment.

\subsection{Rotation measure\label{sec:rm}}

The orientation of polarization vectors is changed in a magnetic plasma
by Faraday rotation that is proportional to the line-of-sight integral
over the density of thermal electrons multiplied by the strength of the
regular field component along the line of sight. Aligned anisotropic fields
do not lead to Faraday rotation.

The rotation measure was calculated with polarization angle maps
which were clipped at $3\sigma$ in polarized emission.
This leads to a maximum uncertainty of the rotation measure of $\sim 16$~rad/m$^{2}$.
Since the 20~cm data suffer from
severe Faraday depolarization, we detect significant polarized
emission at 20~cm only in NGC~4321, NGC~4501, NGC~4396, NGC~4535,
and NGC~4654.
The intensity-weighted mean values of the rotation measure for our
galaxies are presented in Table~\ref{tab:rm}.
Since the rotation
measure is caused by the regular magnetic field component along the
line-of-sight, we expect the rotation measure from an axisymmetric field
to change sign from
one side of the major axis to the other. This is the reason why we
give two values for the NGC~4654 which is the only galaxy where we
can significantly determine rotation measures on both sides of the
minor axis.
\begin{table}
\caption{Polarized intensity weighted mean values of the rotation measure distribution in rad/m$^2$.}
\label{tab:rm}      
\centering                          
\begin{tabular}{l c c c c c c c}        
\hline\hline                 
 & N4321 & N4396 & N4501 & N4535 & N4654 \\    
\hline                        
$<$RM$>$ & -5 & -7 & +10 & 2 & -8\ /\ 2 \\
\hline                                   
\end{tabular}
\end{table}
In general, all face-on galaxies show detectable rotation measures
and we find rotation measures between -7 and
10~rad/m$^2$. However, since the difference in frequency is large
and Faraday depolarization can be strong at 20~cm, one has to
take these measurements with caution. The actual rotation measures
are most probably much larger. For comparison,
We\.zgowiec et al. (2007) report rotation measures between 4.85 and
10.45~GHz obtained from Effelsberg observations of $+14 -
+46$~rad/m$^2$ for NGC~4501 and $-57/+60$~rad/m$^2$ for NGC~4654.
Ultimately, we would like to observe these galaxies at 3~cm to
produce more reliable rotation measure maps. These will help us to
discriminate between the 2 creation scenarios of the polarized
ridges: in the case of shear or compression of a random field, the
resulting ordered field is anisotropic and cannot produce Faraday rotation.
A detectable rotation measure would show regular large-scale fields
and thus help to distinguish them from anisotropic ones. A high
rotation measure in regions of the ridges of enhanced polarized
emission might indicate that a pre-existing large-scale field has
been amplified by compression or shear.

\section{Discussion\label{sec:discussion}}

\subsection{Disk-wide radio halos \label{sec:dwh}}

Only NGC~4402 out of three edge-on galaxies shows a disk-wide radio halo
that is compressed on one side and extends to $\sim 3.7$~kpc above the 
disk plane on the other side (Crowl et al. 2005).
With increasing distance from the galactic disk the total
power surface brightness decreases and the radio continuum spectrum steepens
due to the aging of relativistic electrons generated in the galactic
disk. This halo shows extraplanar polarized emission at the eastern
and western extremities and vertical magnetic field lines in the east,
unlike classical X-structures (Beck 2005 
and references therein, e.g. T\"{u}llmann et al. 2000), 
supporting the scenario of a ram pressure pushed/stripped synchrotron halo.

\begin{figure}
  \resizebox{\hsize}{!}{\includegraphics{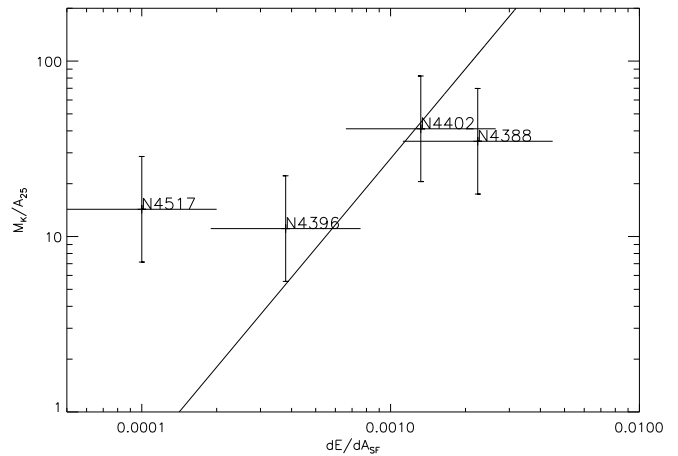}}
  \caption{Mean stellar mass density as a function of the mean
  energy input per unit surface area. The solid line roughly
  indicates the transition zone between galaxies with and without
  radio halos (Dahlem et al. 2006). The error bars represent 
  uncertainties of a factor of 2.}
  \label{fig:radiohalos}
\end{figure}

Dahlem et al. (2006) showed that disk-wide radio halos exist in
galaxies with a low mean mass surface density and a high mean energy
input per unit surface area (Fig.~7 of Dahlem et al. 2006).
To investigate if our edge-on spirals should host a radio halo,
we calculated the mean stellar mass surface density using the
H band magnitudes and optical diameters from Goldmine.
For the mean energy input per unit surface area we use the 20cm flux
densities and radial extents. For NGC~4388 we subtracted the
emission from the active nucleus.
Fig~\ref{fig:radiohalos} shows the
border between galaxies with and without radio halos together with
the data points of our 3 edge-on galaxies. Based on this plot
NGC~4388 and NGC~4402 are expected to show a disk-wide radio halo, but only
NGC~4402 does. A radio halo probably exists in NGC~4388, but
may have been missed due to the higher noise in our maps compared to NGC~4402.
If this is not the case, the pressure of the intracluster medium, 
that confines the radio halo might be higher for NGC~4402 than for NGC~4388.

\subsection{Extraplanar radio continuum emission can extend further
than the H{\sc i} emission}

\begin{figure}
  \resizebox{\hsize}{!}{\includegraphics{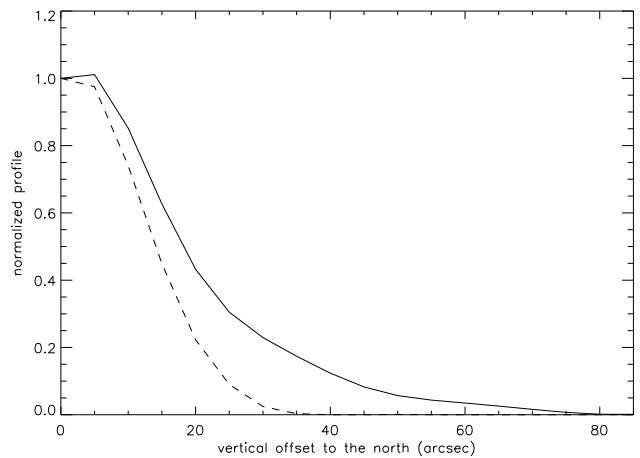}}
  \caption{NGC~4402: vertical profiles to the north of the 20cm continuum 
  (solid line) and H{\sc i} (dashed line) emission distribution
  at a resolution of $20''$.}
  \label{fig:n4402profiles}
\end{figure}

In our sample of 8 Virgo spiral galaxies we have two examples of
asymmetric extraplanar total power emission extending further than the H{\sc i}
emission: NGC~4402, and NGC~4438.
For comparison, in NGC~891 the distribution of neutral hydrogen
has a vertical extent comparable to that of the radio halo (Oosterloo et al. 2007).
This is not the case for NGC~4402 (Fig.~\ref{fig:n4402profiles}) where the radio 
halo is more extended to the north than the H{\sc i} distribution.

The extraplanar radio continuum
emission of NGC~4438, located up to $5$~kpc from the galactic disk, 
is unique in the sense that it has an almost
constant surface brightness and spectral index (Vollmer et al. 2009).\\ 
NGC~4396 and NGC~4402 show extraplanar
6~cm polarized emission in the southwestern (NGC~4396) and northeastern (NGC~4402) 
part of the disk extending $\sim 2.5$~kpc above the galactic plane. 
For NGC~4396 its association to the extraplanar total
power emission is uncertain, because a relatively bright point
source occupies the same region. In NGC~4402 this extraplanar polarized emission
is associated with total power emission from the radio halo.
NGC~4396 thus resembles NGC~4402, but it lacks a 
detectable radio halo, because of its low mean star formation rate (Fig.~\ref{fig:radiohalos}).

The radio continuum emission of the galaxies in our sample and in the
literature (i) is more extended than the extraplanar H$\alpha$ and H{\sc i} emission in
NGC~4402, NGC~4438, and NGC~4569 (Chy\.zy et al. 2006), (ii) follows the H$\alpha$ and H{\sc i} emission 
in NGC~4501, and NGC~4522 (Vollmer et al. 2004), and (iii) is less extended than the H{\sc i} emission in 
NGC~4535 and NGC~4654.
The asymmetric extraplanar radio continuum emission extending further than the H{\sc i} emission
may be due to (i) the ionization of ram pressure stripped H{\sc i},
(ii) more efficient stripping of the cosmic ray gas with its
associated magnetic field, or (iii) nuclear outflows.

\subsection{Ram pressure compression does not change the local total
power emission nor the spectral index}

In all galaxies with asymmetric polarized emission ridges we do not
detect significant enhanced total power emission associated with
these ridges (Sect.~\ref{sec:totalpower}). This means that the local
star formation and the small-scale turbulent magnetic field, which
is sensitive to star formation in unperturbed galaxies, are not
affected by the external ram pressure compression. 
Murphy et al. (2009) came to the same conclusion 
investigating the radio-FIR relation
of perturbed Virgo spiral galaxies based on Spitzer 70~$\mu$m and
VIVA 20~cm data. They found no local but a slight global radio
excess (by a factor of $\sim 2$) in NGC~4330, NGC~4388, and
NGC~4522. We suggest that this excess is due to a slight compression of the
small-scale turbulent magnetic field.
On the other hand, the large-scale magnetic field and thus the 
polarized emission is influenced by the external ram pressure compression.
This compression leads to the observed asymmetric ridges of polarized emission.
Our result, that the small-scale magnetic field is not significantly
compressed whereas the large-scale field is, can be understood in terms of relevant timescales.
Whereas the timescale for ram pressure stripping is several $\sim 10$~Myr 
(see, e.g., Vollmer et al. 2001),
the free fall time of molecular clouds leading to star formation is several Myr 
(see, e.g., Krumholz \& Tan 2007). Thus the dynamics of the mainly
molecular gas in the star forming spiral arms are decoupled from the
overall large-scale (~1~kpc) motions. 

In the same line, the spectral index between 20 and 6~cm in our sample galaxies 
only depends on the local star formation rate, as it is the case for unperturbed galaxies
(Sukumar et al. 1987, Berkhuijsen et al. 2003, Tabatabaei et al. 2007a/b, Heesen et al. 2009), 
Sect.~\ref{sec:spectralindex}). 
This is contrary to what we found in NGC~4522
(Vollmer et al. 2004) where the radio continuum spectrum flattens toward the edge of the
disk where the gas is compressed. Since NGC~4522 is close to peak ram pressure (Vollmer et al. 2006),
a flattening of the spectrum in the compressed region might
only occur in galaxies undergoing very strong ram pressure stripping.
Unfortunately, the quality of the 20cm data of the second galaxy in our sample 
which is close to peak ram pressure, NGC~4438, does not permit
to determine the spectral index within the galactic disk.

We thus conclude
that ram pressure by the intracluster medium leads to gas compression on
large scales ($\ga 1$~kpc; see also Vollmer et al. 2008 for a
detailed discussion of NGC~4501) without a significant enhancement of
the star formation and the associated small-scale turbulent magnetic field.
This is consistent with the results of Koopmann \& Kenney (2004)
who found that only 2\% of their 52 Virgo cluster spiral galaxies have
an enhanced star formation distribution. Only one galaxy shows a truncated and
enhanced star formation distribution. NGC~4321, NGC~4501, and NGC~4522 are classified
as having truncated normal star formation distributions, NGC~4535 and NGC~4654
as having normal star formation distributions.

\subsection{The degree of polarization gives additional information
on the nature of the interaction}

We suggest that the vertically and azimuthally asymmetric increase 
of the degree of polarization (Sect.~\ref{sec:polarizedemission}) is due 
to ram pressure stripping and thus indicates the ram pressure wind direction 
or the direction of the galaxy's motion in a static intracluster medium. This is
consistent with Murphy et al. (2009) who found radio deficits with
respect to the FIR surface brightness in the regions where we
observe an enhanced degree of polarization in NGC~4402 and NGC~4522 
(Vollmer et al. 2004).
The absence of a local radio-deficit
region at the southern edge of NGC~4388's disk is probably due to
the radio continuum emission from the nuclear outflow (Veilleux et al. 1999).

In NGC~4388 (Vollmer \& Huchtmeier 2003), NGC~4402 (Crowl et al. 2005), 
and NGC~4501 (Vollmer et al. 2008) the degree of polarization is highest
in the direction of the ram pressure wind. In unperturbed spiral galaxies 
galactic rotation leads to an azimuthally symmetric large-scale magnetic field.
Due to beam depolarization the degree of polarization decreases towards the edge 
of the galactic disk in highly inclined galaxies.
We interpret the increase of the degree of
polarization along the major axis in NGC~4396 and NGC~4402 as ionized ISM
that deviates significantly from galactic rotation. 
In NGC~4402 the western maximum of polarized emission thus traces ram pressure
stripped ionized gas with its associated magnetic field. This is consistent with
the ram pressure scenario of Crowl et al. (2005) where the ram pressure wind comes
from the southeast. By extrapolating this interpretation to NGC~4396, we suggest that
the ram pressure wind direction is south.

\section{Conclusions\label{sec:conclusions}}

Deep VLA 20 and 6~cm radio continuum data including polarization of
a sample of 8 Virgo spiral galaxies are presented and combined with
optical DSS, VIVA H{\sc i} (Chung et al. 2009), and Goldmine
H$\alpha$ data. We study the spatial distributions of the spectral
index, the degree of polarization, and the rotation measure and
derive the following conclusions:
\begin{enumerate}
\item
Ram pressure leads to sharp edges of the total power distribution
on one side of the galactic disk (NGC~4501, NGC~4654). The radio 
continuum edge coincides with the H{\sc i} edge.
\item
In edge-on galaxies, the extraplanar radio continuum emission (total
power and polarized intensity) can extend further than the H{\sc i}
emission. The most prominent example is NGC~4438. 
\item
In the case of edge-on galaxies, we find azimuthally asymmetric
distributions of the degree of polarization (NGC~4388, NGC~4396, NGC~4402).
This asymmetry gives important information on the ram
pressure wind direction.
\item
Ram pressure does not alter the local total power emission nor the
spectral index. Only very strong ram pressure might lead to a flattening of
the radio continuum spectrum. 
This means that the local star formation and the
small-scale turbulent magnetic field, which is sensitive to star
formation in unperturbed galaxies, are not influenced by external ram
pressure compression. We thus conclude that ram pressure by the
intracluster medium leads to compression on large-scales ($\ge
1$~kpc). In addition, the absence of enhanced total power
emission also implies that star formation is not significantly
enhanced in the compressed regions.
\end{enumerate}
Ultimately, we would like to observe these galaxies at 3~cm in
polarized emission to produce reliable rotation measure maps between
3 and 6~cm. Furthermore, single-dish observations at 3~cm and 6~cm
are needed to fill the missing spacings of interferometric observations.
These will help us to discriminate between the 2
creation scenarios of the polarized ridges: in the case of shear or
compression of a random field, the resulting ordered field is
anisotropic and has no Faraday rotation; whereas in the case of a
large-scale magnetic field amplification by compression or shear
Faraday rotation is expected to be high.

\begin{acknowledgements}
This research has made use of the GOLD Mine Database. This work was
supported by the Polish-French (ASTRO-LEA-PF) cooperation program
and by the Polish Ministry of Sciences and Higher Education grant
3033/B/H03/2008/35. The authors would like to thank E.M.~Berkhuijsen
for careful reading of the manuscript.
\end{acknowledgements}

\end{document}